\pdfoutput=1
\documentclass[twoside,twocolumn,9pt]{article}
\usepackage{extsizes}
\usepackage[super,sort&compress,comma]{natbib} 
\usepackage[version=3]{mhchem}
\usepackage[left=1.5cm, right=1.5cm, top=1.785cm, bottom=2.0cm]{geometry}
\usepackage{balance}
\usepackage{widetext}
\usepackage{sectsty}
\usepackage{graphicx} 
\usepackage{lastpage}
\usepackage[format=plain,justification=raggedright,singlelinecheck=false,font={stretch=1.125,small,sf},labelfont=bf,labelsep=space]{caption}

\usepackage{float}
\usepackage{fancyhdr}
\usepackage{fnpos}
\usepackage[english]{babel}
\usepackage{array}
\usepackage{droidsans}
\usepackage{charter}
\usepackage[T1]{fontenc}
\usepackage[usenames,dvipsnames]{xcolor}
\usepackage{setspace}
\usepackage[compact]{titlesec}

\definecolor{cream}{RGB}{222,217,201}

\usepackage{cmap}
\usepackage{fixmath}

\usepackage[slantedGreek]{mathptmx}
\usepackage{times}
\usepackage{fixmath}
\usepackage[Symbolsmallscale]{upgreek}

\usepackage{mciteplus}
\usepackage{mleftright}

\usepackage{booktabs}

\usepackage[pdftex, bookmarks, pdffitwindow=false, pdfstartview=FitH, pdfdisplaydoctitle, colorlinks, plainpages=false,pdftitle={Effects of co-ordination number on the nucleation behaviour in many-component self-assembly}, pdfauthor={Aleks Reinhardt, Chon Pan Ho, Daan Frenkel}, pdfpagelabels, hypertexnames, citecolor={blue},linkcolor={red}, urlcolor={violet}, pdflang={en}, hyperfootnotes=false, breaklinks]{hyperref}

\usepackage{graphicx}
\usepackage{amsmath}

\usepackage[dvipsnames]{xcolor}

\usepackage[stretch=15,shrink=15,step=1]{microtype}
\usepackage[obeyall]{siunitx}

\begin{document}

\pagestyle{fancy}
\thispagestyle{plain}
\fancypagestyle{plain}{

\fancyhead[C]{\includegraphics[width=18.5cm]{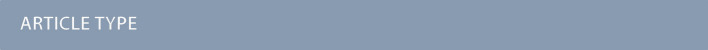}}
\fancyhead[L]{\hspace{0cm}\vspace{1.5cm}}
\fancyhead[R]{\hspace{0cm}\vspace{1.7cm}\includegraphics[height=55pt]{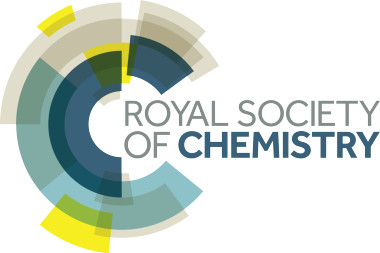}}
\renewcommand{\headrulewidth}{0pt}
}

\makeFNbottom
\makeatletter
\renewcommand\LARGE{\@setfontsize\LARGE{15pt}{17}}
\renewcommand\Large{\@setfontsize\Large{12pt}{14}}
\renewcommand\large{\@setfontsize\large{10pt}{12}}
\renewcommand\footnotesize{\@setfontsize\footnotesize{7pt}{10}}
\makeatother

\renewcommand{\thefootnote}{\fnsymbol{footnote}}
\renewcommand\footnoterule{\vspace*{1pt}%
\color{cream}\hrule width 3.5in height 0.4pt \color{black}\vspace*{5pt}} 
\setcounter{secnumdepth}{5}

\makeatletter 
\renewcommand\@biblabel[1]{#1}            
\renewcommand\@makefntext[1]%
{\noindent\makebox[0pt][r]{\@thefnmark\,}#1}
\makeatother 
\renewcommand{\figurename}{\small{Fig.}~}
\sectionfont{\sffamily\Large}
\subsectionfont{\normalsize}
\subsubsectionfont{\bf}
\setstretch{1.125} 
\setlength{\skip\footins}{0.8cm}
\setlength{\footnotesep}{0.25cm}
\setlength{\jot}{10pt}
\titlespacing*{\section}{0pt}{4pt}{4pt}
\titlespacing*{\subsection}{0pt}{15pt}{1pt}

\fancyfoot{}
\fancyfoot[LO,RE]{\vspace{-7.1pt}\includegraphics[height=9pt]{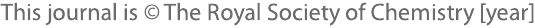}}
\fancyfoot[CO]{}
\fancyfoot[CE]{\vspace{-7.2pt}\hspace{-14.2cm}\includegraphics{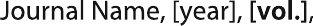}}
\fancyfoot[RO]{\footnotesize{\sffamily{1--\pageref{LastPage} ~\textbar  \hspace{2pt}\thepage}}}
\fancyfoot[LE]{\footnotesize{\sffamily{\thepage~\textbar\hspace{3.45cm} 1--\pageref{LastPage}}}}
\fancyhead{}
\renewcommand{\headrulewidth}{0pt} 
\renewcommand{\footrulewidth}{0pt}
\setlength{\arrayrulewidth}{1pt}
\setlength{\columnsep}{6.5mm}
\setlength\bibsep{1pt}

\makeatletter 
\newlength{\figrulesep} 
\setlength{\figrulesep}{0.5\textfloatsep} 

\newcommand{\topfigrule}{\vspace*{-1pt}%
\noindent{\color{cream}\rule[-\figrulesep]{\columnwidth}{1.5pt}} }

\newcommand{\botfigrule}{\vspace*{-2pt}%
\noindent{\color{cream}\rule[\figrulesep]{\columnwidth}{1.5pt}} }

\newcommand{\dblfigrule}{\vspace*{-1pt}%
\noindent{\color{cream}\rule[-\figrulesep]{\textwidth}{1.5pt}} }

\makeatother

\twocolumn[
  \begin{@twocolumnfalse}
\vspace{3cm}
\sffamily
\begin{tabular}{m{4.5cm} p{13.5cm} }

DOI:~\href{http://dx.doi.org/10.1039/C5FD00135H}{10.1039/C5FD00135H} & \noindent\LARGE{\textbf{Effects of co-ordination number on the nucleation behaviour in many-component self-assembly}} \\
\vspace{0.3cm} & \vspace{0.3cm} \\
 & \noindent\large{Aleks Reinhardt$^{a}$, Chon Pan Ho$^{a}$ and Daan Frenkel$^{a}$} \\

 & \noindent\normalsize{We report canonical and grand-canonical lattice Monte Carlo simulations of the self-assembly of addressable structures comprising hundreds of distinct component types. The nucleation behaviour, in the form of free-energy barriers to nucleation, changes significantly as the co-ordination number of the building blocks is changed from 4 to 8 to 12. Unlike tetrahedral structures -- which roughly correspond to DNA bricks that have been studied in experiment -- the shapes of the free-energy barriers of higher co-ordination structures depend strongly on the supersaturation, and such structures require a very significant driving force for structure growth before nucleation becomes thermally accessible. Although growth at high supersaturation results in more defects during self-assembly, we show that high co-ordination number structures can still be assembled successfully in computer simulations and that they exhibit self-assembly behaviour analogous to DNA bricks. In particular, the self-assembly remains modular, enabling in principle a wide variety of nanostructures to be assembled, with a greater spatial resolution than is possible in low co-ordination structures.} \\

\end{tabular}

 \end{@twocolumnfalse} \vspace{0.6cm}

  ]

\renewcommand*\rmdefault{bch}\normalfont\upshape
\rmfamily
\section*{}
\vspace{-1cm}


\footnotetext{\textit{$^{a}$~Department of Chemistry, University of Cambridge, Lensfield Road, Cambridge, CB2~1EW, United Kingdom.}}






\section{Introduction}
Materials that can be formed by self-assembly have over time become increasingly more complex;\cite{Whitelam2015} furthermore, in the last few years, the field has seen something of an explosion in the number of self-assembling materials which exhibit not only structural complexity, but which are `addressably' complex,\cite{Frenkel2015} in the sense that the individual building blocks making up these structures are all distinct. Such self-assembled materials are not only interesting from the point of view of fundamental science, but are thought to hold considerable promise for applications in many aspects of nanotechnology,\cite{Linko2013, *Sobczak2015} especially since the addressable nature of the building blocks should allow the structures to be functionalised with sub-nanometre-scale resolution.

Recent experiments have demonstrated that it is possible to assemble structures comprising thousands of distinct modular building blocks into well-formed target structures by making use of single-stranded DNA molecules -- termed `DNA bricks' -- designed to have an obligate set of hybridisation partners, in the sense that those parts of the DNA molecules that are designed to be bonded in the target structure have complementary sequences.\cite{Ke2012, Wei2012, Zhang2013c, Wei2013, Ke2014} In the past few years, several theoretical and computational studies have also been undertaken, probing the intriguing self-assembly behaviour exhibited by such systems.\cite{Halverson2013, Reinhardt2014, Hedges2014, Zenk2014, Zeravcic2014, Murugan2015, Jacobs2015, Jacobs2015b, Whitelam2015b, Madge2015}

We have previously shown that the DNA brick self-assembly is made possible by the interplay between self-assembly and growth. In particular, as a system of DNA bricks is cooled, at some temperature the free-energy barrier to nucleation becomes small enough that nucleation can occur, but nucleation events remain sufficiently rare that any clusters that do form do not interact significantly with one another, and monomers are not initially significantly depleted,\cite{Reinhardt2014, Jacobs2015b} which enables these clusters to grow in an essentially error-free manner as the temperature is decreased. However, such behaviour only occurs over a very narrow window of temperatures: if the experiment is performed at a low temperature from the outset, misassembled aggregates dominate instead.\cite{Reinhardt2014} Nucleation thus plays an important role in enabling structures of this type to self-assemble successfully.

Whilst DNA bricks have been shown to self-assemble reliably, our previous theoretical work has indicated that the co-ordination number of the particles that form self-assembling structures determines their nucleation behaviour in both two and three dimensions.\cite{Jacobs2015b} In particular, the larger the co-ordination number, the more classical-looking the free-energy barrier to nucleation becomes. Yet one of the key aspects that seemed to enable the lower co-ordinate structures to form successfully was the non-classical nucleation barrier. Specifically, for tetrahedrally co-ordinated building blocks, the critical cluster size was found to be largely insensitive to the nature of the target structure, and the nucleation barrier is significant but surmountable at the point at which a large, nearly fully assembled cluster of the designed structure is thermodynamically stable. By contrast -- and in agreement with the predictions of classical nucleation theory -- for higher co-ordination number structures, the free energy barrier to nucleation changes with temperature and is considerably larger than for tetrahedral structures at the same supersaturation.\cite{Jacobs2015b} This suggests that, in order to overcome the free-energy barrier to nucleation, the driving force for growth must increase, for example by increasing the monomer concentration, reducing the temperature or increasing the bond strengths by choosing a different set of DNA sequences. Such approaches, however, would make the competing structures in which monomers have not assembled as designed ever more stable, and our previous theoretical work thus suggests that, as the co-ordination number increases, the structures should become more and more difficult to form.

However, in order to create more varied target structures in an addressable way, we may well need to move to a system with a higher co-ordination number, as this should in principle allow us to construct structures with finer small-scale features due to the considerably greater spatial resolution of the system than we can achieve using tetrahedrally co-ordinated particles. Moreover, a greater degree of bonding can help to stabilise such structures, which may also be important in practical applications.

Although DNA bricks are tetrahedrally co-ordinated,\cite{Ke2012} there are many possible ways in which addressable structures with higher co-ordination number might be experimentally realised. For example, one can envisage that colloidal particles with carefully positioned DNA strands grafted onto the particle in the correct geometry might be possible to assemble in the near future, perhaps similar to the experiments of Wang~\textit{et~al.}\cite{Wang2012} or Lu~\textit{et~al.},\cite{Lu2015} but with each particle functionalised with a unique set of DNA strands.  Alternatively, DNA Holliday junctions and multi-arm motifs\cite{Holliday1964, *Lilley2000, *Zhang2008} can be synthesised to correspond to high co-ordination number structures. Of course in practice, producing structures of this type in experiments may be non-trivial  because, in our examples, each colloidal particle would have to be created with a unique set of grafted DNA strands, and each DNA junction with a different sequence would have to be pre-assembled. It is therefore important that future experiments focus on strategies that are likely to be successful. It is with this in mind that we have carried out the simulations presented here: if structures of this kind cannot be assembled on a computer with a toy model, then it may be risky to attempt to do so experimentally in the light of the significant cost and effort likely to be involved. 

\section{Simulation methods}

We perform canonical ensemble simulations on a lattice, with periodic boundary conditions, using a Metropolis Monte Carlo\cite{Metropolis1953} scheme. To determine the free-energy barriers as a function of the size of the largest crystalline cluster in the system, we use umbrella sampling with adaptive weights\cite{Torrie1977, *Mezei1987} in a time-step separated\cite{Hetenyi2002} Monte Carlo scheme. We use `virtual moves'\cite{Whitelam2007, *Whitelam2008} to allow for realistic dynamics of cluster motion. In our simulations, clusters are randomly translated or rotated on a lattice, with 24 permissible orientations per particle, corresponding to all the possible neighbour interactions on a cubic lattice.\cite{Reinhardt2014}

Each particle in the system is hard in the sense that dual occupancy of lattice sites is not permitted, and each particle has $n$ `patches', where $n$ is the co-ordination number. Every patch is assigned a DNA sequence such that, in the fully assembled target structure, adjacent patches have a complementary sequence, but otherwise these sequences are randomly assigned (subject to the rules identified by Wei~\textit{et~al.}\cite{Wei2012}).\footnote{It is by no means essential for particles in our system to interact via DNA hybridisation; it is sufficient that they have specific, designed interactions. In practice, however, we anticipate at this stage that DNA is the most likely candidate for an experimental realisation of such systems, and we have chosen to parameterise our model accordingly.} Particles that are adjacent to each other interact with a slightly repulsive energy $\varepsilon_\text{init}/k_\text{B}=\SI{100}{\kelvin}$,\cite{Reinhardt2014} to which we add the hybridisation free energy of the longest complementary sequence match between the nearest pair of patches, calculated using a standard thermodynamic model.\cite{SantaLucia2004} As in the experimental work of Ke~\textit{et al.},\cite{Ke2012} the outermost particles in the target structure are assigned a poly-T sequence to minimise any misbonding.

Particles which have 4 and 8 patches have a minimum interparticle distance of $a\sqrt{3}$, where $a$ is the lattice parameter,\cite{Reinhardt2014} whilst particles with 12 patches have a minimum interparticle distance of $a\sqrt{2}$ to be able to accommodate the additional neighbours. This means that the effective densities are not strictly comparable, as the lower co-ordination structures have a greater excluded volume.

In grand canonical simulations, we introduce particle addition and removal moves in addition to the canonical (virtual move) translations and rotations. Particles to be added or removed are chosen at random. A particle addition move in which a particle of type $i$ has been placed at a random position and with a random orientation in the simulation box is accepted with probability\cite{Frenkel2002}\footnote{The de Broglie thermal wavelength is subsumed into the chemical potential, and cancels out in the case of an ideal chemical potential. For convenience, we have therefore set it to unity.}
\begin{equation}
P_\text{acc}^\text{add} = \min\mleft[1,\,\frac{V z_i}{N_i+1} \exp(-\upDelta E /k_\text{B}T)\mright],
\end{equation}
where $V$ is the volume of the simulation box, $N_i$ is the current number of particles of type $i$ in the system, $\upDelta E$ is the trial change in the system's potential energy, and $z_i$ is the fugacity of particles of type $i$, i.e.~$z_i=\exp(\mu_i/k_\text{B}T)$, where $\mu_i$ is the particle's chemical potential. The ideal chemical potential is given by $\mu_\text{id}=k_\text{B}T \ln \rho$, where $\rho$ is the number density; in the absence of interactions, the fugacity thus determines the target number density. An analogous acceptance probability holds for particle removals,
\begin{equation}
 P_\text{acc}^\text{rem} =  \min\mleft[1,\,\frac{N_i}{V z_i N_\text{types}} \exp(\upDelta E /k_\text{B}T)\mright],
\end{equation}
where $N_\text{types}$ is the number of types of particle in the system. This accounts for the fact that when we add a particle, we choose its type uniformly at random, whereas when we remove a particle, we choose the particle at random: in order to obey detailed balance, we must account for the probability of choosing a particle of each type.

\begin{figure*}[tbp]
\centering
\includegraphics{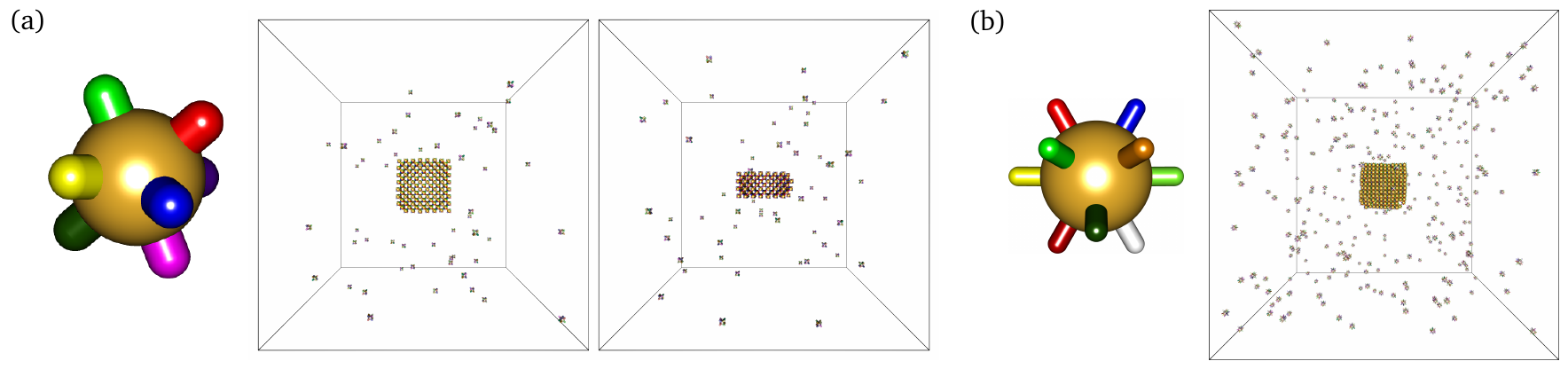}
\caption{A single monomer and snapshots towards the end of the nucleation process of some structures assembled from a vapour of monomers for the (a) 8 (\SI{332}{\kelvin}) and (b) 12 (\SI{345}{\kelvin}) co-ordinate monomers. The target structures in each case were simple rectangular parallelepipeds. In the simulation snapshots, correctly bonded clusters are shown in the same colour, but each particle, and each patch, is in fact distinct.}\label{fig-coordNum-structures}
\end{figure*}

\section{Results and discussion}
We have previously considered tetrahedral co-ordination, as applicable to DNA bricks. Here, we investigate the self-assembly behaviour of structures with a co-ordination number of 8 (giving bcc-like target structures) and 12 (giving fcc-like target structures). The corresponding building blocks and sample target structures are shown in Fig.~\ref{fig-coordNum-structures}. The sequences associated with each patch for the structures we have studied are provided as supporting data.\footnote{It is important to bear in mind that, if these sequences are chosen randomly, the temperatures at which nucleation and growth occur can change by a few degrees in identical conditions.\cite{Reinhardt2014} The temperatures we quote in the text refer to these specific DNA sequences. While the numerical values change with sequence choice, the qualitative behaviour does not.}  In the simulations reported here, the numbers of distinct particles in the target structures was 396 for the 4-, 403 for the 8- and 256 for the 12-co-ordinated structures.

\begin{figure}[tbp]
\centering
\includegraphics{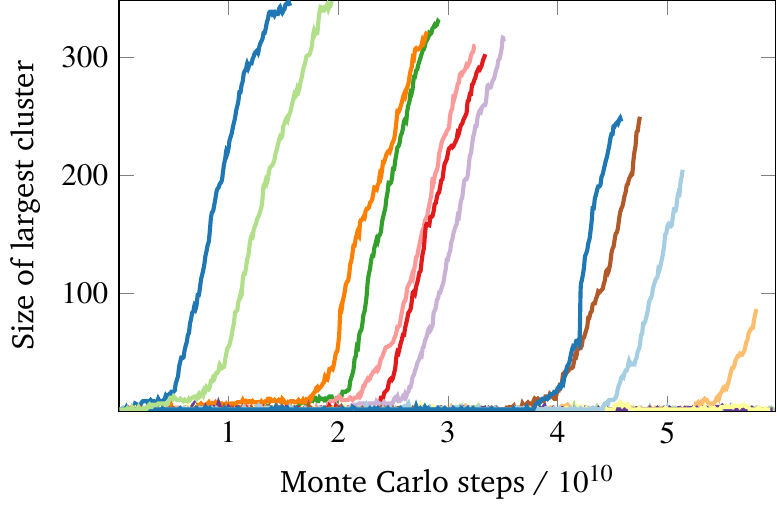}
\caption{The size of the largest cluster in the system as a function of Monte Carlo time for a canonical simulation with a total of 403 distinct particles with a co-ordination number of 8. $T=\SI{332}{\kelvin}$, $\rho=1/(62a)^3$. The different colours correspond to individual Monte Carlo trajectories started from an equilibrated vapour of monomers. These trajectories were run for a fixed real-time; since virtual moves make simulations of larger clusters slower, simulations in which nucleation occurred later could run for a larger number of Monte Carlo steps.}\label{fig-bcc-clusterSize-vs-time}
\end{figure}

\begin{figure}[t!]
\centering
\includegraphics{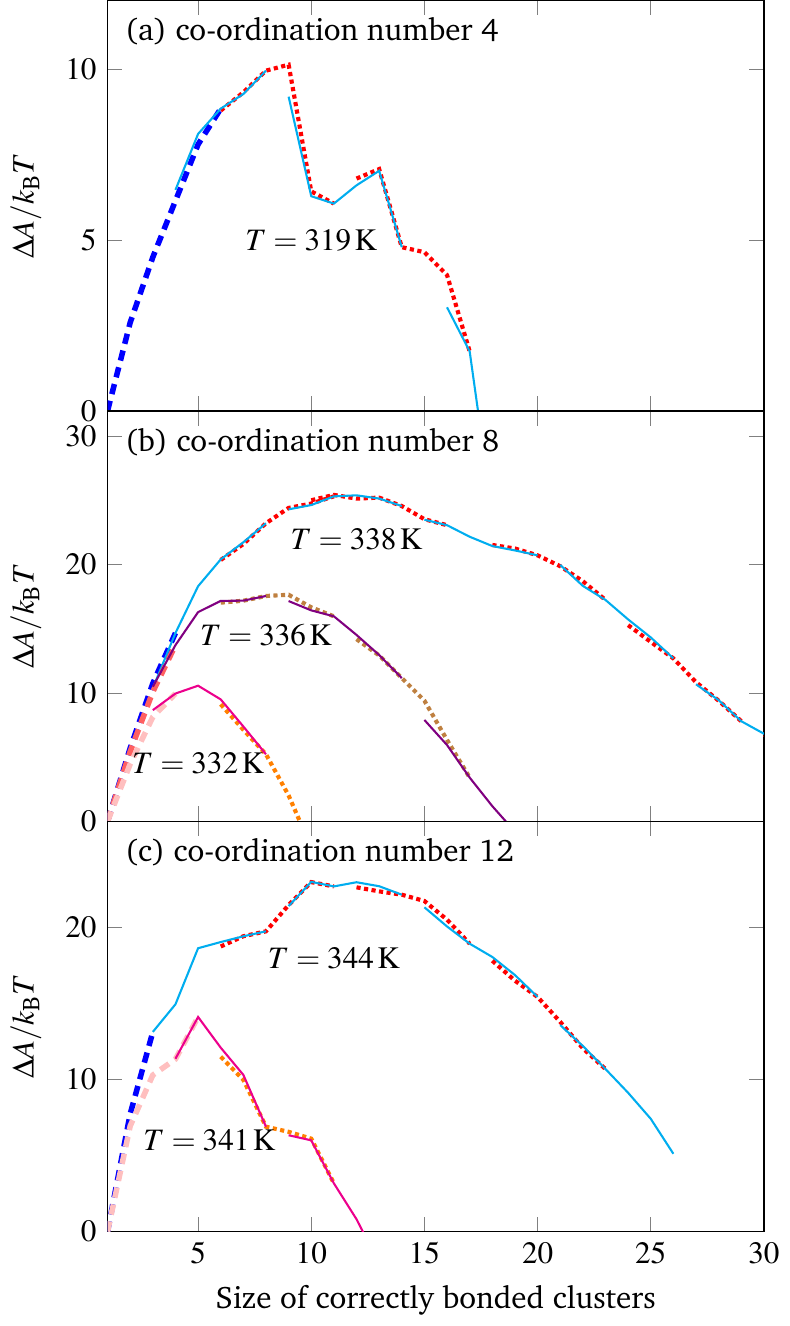}
\caption{The free-energy profile for cluster growth of particles with a co-ordination number of (a) 4, (b) 8 and (c) 12. Simulation results from different umbrella sampling windows are depicted in alternating styles to show their overlap. The thick dashed line corresponds to brute-force simulations.\cite{Reinhardt2014} In each case, there was one copy of each particle in the simulation box of dimensions $62a\times 62a \times 62a$.}\label{fig-coordNum-freeenergybarrier}
\end{figure}

Contrary to expectations,\cite{Jacobs2015b} brute force simulations starting from a vapour of one copy of each of the monomers required to assemble a single target structure can, within a narrow temperature window, result in the successful self-assembly of the target structures shown in Fig.~\ref{fig-coordNum-structures}. Moreover, as evident from Fig.~\ref{fig-bcc-clusterSize-vs-time}, this process is stochastic: under identical thermodynamic conditions, systems can exhibit drastically different lag times before any significant growth occurs. This is indicative of the presence of a free-energy barrier to nucleation, whereby a cluster of a sufficient size must form spontaneously before further growth is thermodynamically favoured. Since monomers coming together to form such a cluster lose a significant amount of translational and orientational entropy, this happens infrequently: there is a free-energy barrier associated with nucleation. Using umbrella sampling, we have calculated this free-energy barrier\footnote{The order parameter used as a collective variable, i.e.~the number of particles in the largest cluster, is a convenient choice consistent with classical nucleation theory. However, because each particle is different in these simulations, any \emph{particular} cluster that forms can behave rather differently from this averaged behaviour. This is especially important if the cluster under consideration forms near a face or an edge of the target structure, where the average environments are different from those at the centre of the structure.} for the two target structures shown in Fig.~\ref{fig-coordNum-structures} at a number of temperatures, as shown in Fig.~\ref{fig-coordNum-freeenergybarrier}, where we also show a free-energy barrier for a reference tetrahedral system. Of course higher co-ordination structures are more stable at higher temperatures, since such structures entail many more bonds, and so the temperature scale at which nucleation occurs depends on the co-ordination number. Figures \ref{fig-bcc-clusterSize-vs-time} and \ref{fig-coordNum-freeenergybarrier} indicate that the process is indeed nucleation-initiated for both the 8- and 12-co-ordinated target structures. However, whilst the process remains nucleation-initiated, there are significant differences in the systems' behaviour relative to the self-assembly of tetrahedral particles.

In particular, tetrahedrally co-ordinated structures, which include the experimentally studied DNA bricks, exhibit a free-energy barrier with a distinct jagged appearance. This is not an artefact of the simulation technique used or a lack of equilibration, but rather reflects the fact that as clusters grow, there is a competition between the entropy loss associated with monomers losing their translational and vibrational degrees of freedom when they are attached to a larger cluster on the one hand, and on the other the energy gain associated with the formation of `designed' interactions, which are, by construction, highly favourable. Tetrahedral structures grow in a very predictable fashion, with steps at which clusters can form closed cycles, for which the entropic penalty is compensated by not one, but two designed bonds forming, having a considerably lower free energy than other steps do.\cite{Reinhardt2014, Jacobs2015, Jacobs2015b} The critical cluster for tetrahedrally co-ordinated structures is typically bicylic or tricyclic (adamantane-like) with a single particle missing,\cite{Reinhardt2014, Jacobs2015, Jacobs2015b} i.e.~the size of the critical cluster is typically 8 or 9, and this cluster size appears to be essentially temperature independent in the regime where nucleation can occur. By contrast, the free-energy profiles shown in Fig.~\ref{fig-coordNum-freeenergybarrier}, in agreement with our theoretical prediction,\cite{Jacobs2015b} are considerably smoother, and the temperature greatly affects the size of the critical cluster. The reason for this behaviour is that there are a considerably larger number of possibilities of forming different clusters comprising the same number of building blocks;\cite{Jacobs2015b} this makes the nucleation considerably more classical, affecting both the smoothness and the dependence of the critical nucleus size on temperature.

However, despite this quite different behaviour at high temperatures, the systems behave in a less divergent manner at temperatures where the nucleation barrier is sufficiently small compared to thermal energy that nucleation can reasonably be expected to occur. The degree of supercooling required in order to observe a nucleation event is not significantly different amongst the structures we have studied: if we deem the temperature at which a pre-formed target structure fully `melts' to be an effective melting point, nucleation becomes sufficiently fast to observe in brute-force simulations at a supercooling of approximately \SI{2}{\percent} for all target structures considered. The point at which mass aggregation occurs is also similar, at a roughly \SI{4}{\percent} supercooling. These results indicate that a more optimistic view of the possibility of assembling high co-ordination number structures is perhaps warranted.

Nevertheless, one difference in the behaviour observed is noteworthy. At temperatures at which there is a reasonably small free-energy barrier to nucleation, the driving force for growth is considerably larger for higher co-ordination number structures. One proxy for this is the gradient of the free-energy profile at post-critical cluster sizes: this gradient has roughly the same value ($\sim${}$-1.1\,k_\text{B}T$ per particle) in the tetrahedral case where the critical free-energy barrier height is approximately $10\,k_\text{B}T$, and in the 8-co-ordinate structure at \SI{338}{\kelvin} with a critical free-energy barrier height of $25\,k_\text{B}T$. As the temperature is decreased, the effective supersaturation increases: at \SI{332}{\kelvin}, the large-cluster gradient of the free energy is already $-5\,k_\text{B}T$ per particle. This means that the conditions in which the 8- and 12-co-ordinate structures grow are considerably harsher than in the tetrahedral case, which is likely to lead to more mistakes during assembly.\cite{Hedges2014}

In simulations where  only one particle of each component is present, the increased supersaturation may not interfere with correct self-assembly, since competing structures are less likely to occur. Of course, in experiments, many copies of each building block are present. 
To investigate whether higher co-ordination number structures can still form in circumstances where competition from additional monomers and clusters is possible, we have also simulated the self-assembly process in the grand canonical ensemble. We have run simulations at a fugacity corresponding to the same ideal number density as in the canonical simulations, starting from an empty simulation box of various volumes, and we observe successful self-assembly to completion at a number of temperatures for both the 8- and the 12-co-ordinate structures.\footnote{In the grand ensemble, the stability of the target structure at temperatures at which nucleation occurs changes with the co-ordination number: for the tetrahedral structures, partially formed structures dominate, whilst for high co-ordination structures, essentially fully formed structures result at the end of the self-assembly process.} Correctly assembled clusters grow one-by-one in such simulations: at sufficiently high temperatures, nucleation remains a rare event and the clusters grow essentially to completion before additional clusters nucleate. This observation supports the conclusion from canonical simulations that nucleation helps to prevent cluster interactions. These grand-canonical simulations also confirm that the lack of competition from monomers and clusters in solution is not the principal reason why self-assembly can succeed in the canonical ensemble, and the self-assembly process is surprisingly robust.

Despite this apparent success, the prediction that the greater supersaturation leads to more defects does  hold. If we compare the largest assembled structures in the grand ensemble at the highest temperature at which nucleation was found to occur for co-ordination numbers of 4 and 12 (\SI{319}{\kelvin} and \SI{338}{\kelvin}, respectively), the high co-ordination number structures typically have one or two incorrect particles embedded in the structure, and one or two vacancies, whilst the tetrahedral structure is entirely error free. The error rates would, moreover, be expected to be higher still if we implemented a `kinetic constraint' to prevent the change of state for any particle wholly within the solid structure to account for the relative slowness of relaxation dynamics within a solid structure:\cite{Mannige2015} this would, in particular, prevent vacancies from being filled when the rest of the structure has already formed around them. While the number of defects in absolute terms is not large even for the high co-ordination number structures, it is worth bearing in mind that incorrect particles on the surface of the cluster can lead to additional undesired clustering as the temperature is lowered and the clusters allowed to undergo diffusion for long periods of time.

\begin{figure}[tbp]
\centering
\includegraphics{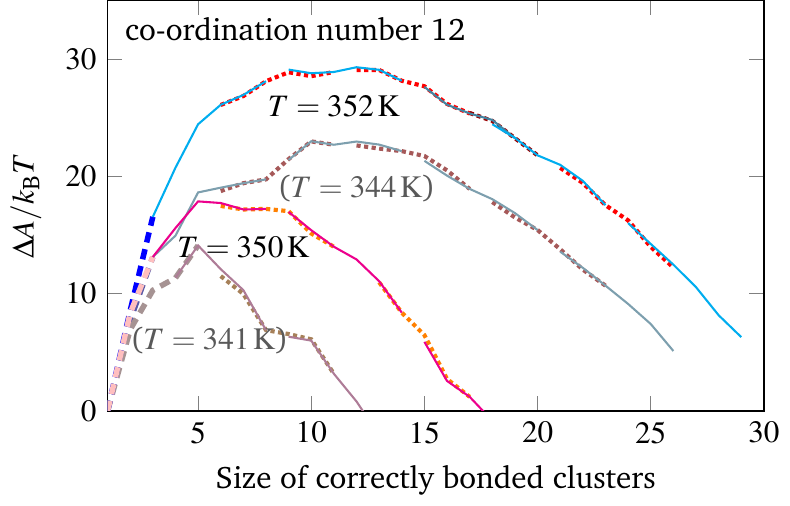}
\caption{The free-energy profile for cluster growth of particles with a co-ordination number of 12 with a GC content of \SI{68}{\percent}. The free-energy profiles of Fig.~\ref{fig-coordNum-freeenergybarrier} are reproduced in a greyed-out hue.}\label{fig-coordNum-freeenergybarrier-ATcontent}
\end{figure}

One way in which the driving force for nucleation can be changed is by strengthening or weakening the average bond energy between particles. When using DNA bases, this can be achieved by varying the proportion of G and C bases at the expense of A and T: the larger the GC content, the stronger on average the hybridisation between two complementary strands will be.\cite{SantaLucia2004} We have therefore simulated the self-assembly of the same target structures, but with differently chosen patch sequences. These are still chosen randomly, but with an appropriate bias towards either GC or AT base pairs.\footnote{Terminal poly-T sequences are ignored in the GC content calculation.} Because the DNA hybridisation free energy itself depends strongly on the temperature, changing the bond strengths in this way is not equivalent to simply shifting the temperature scale. We show in Fig.~\ref{fig-coordNum-freeenergybarrier-ATcontent} some additional free-energy barriers calculated for a system with stronger average interactions. Whilst the basic behaviour remains unchanged, the different temperature at which nucleation becomes feasible does affect the driving force for growth and thus the likelihood of defects occurring during the process. For example, if we compare the curves corresponding to $T=\SI{352}{\kelvin}$ and $T=\SI{344}{\kelvin}$ in Fig.~\ref{fig-coordNum-freeenergybarrier-ATcontent}, the system with weaker bonds has a less negative large-cluster gradient of the free energy as a function of the largest cluster size ($\sim${}$-1.1\,k_\text{B}T$ per particle compared to $\sim${}$-1.5\,k_\text{B}T$ per particle) and thus has a weaker driving force for growth, even though the nucleation free-energy barrier is considerably smaller ($23\,k_\text{B}T$ compared to $29.5\,k_\text{B}T$). Moreover, the system with stronger bonds appears to grow with more defects in a grand canonical simulation, with typically three or four incorrect particles bonded in the final structure. A judicious choice of DNA sequences can thus significantly affect the probability that high co-ordination number structures in particular can grow in a reasonably error-free manner.

\begin{figure}[b!]
\centering
\includegraphics{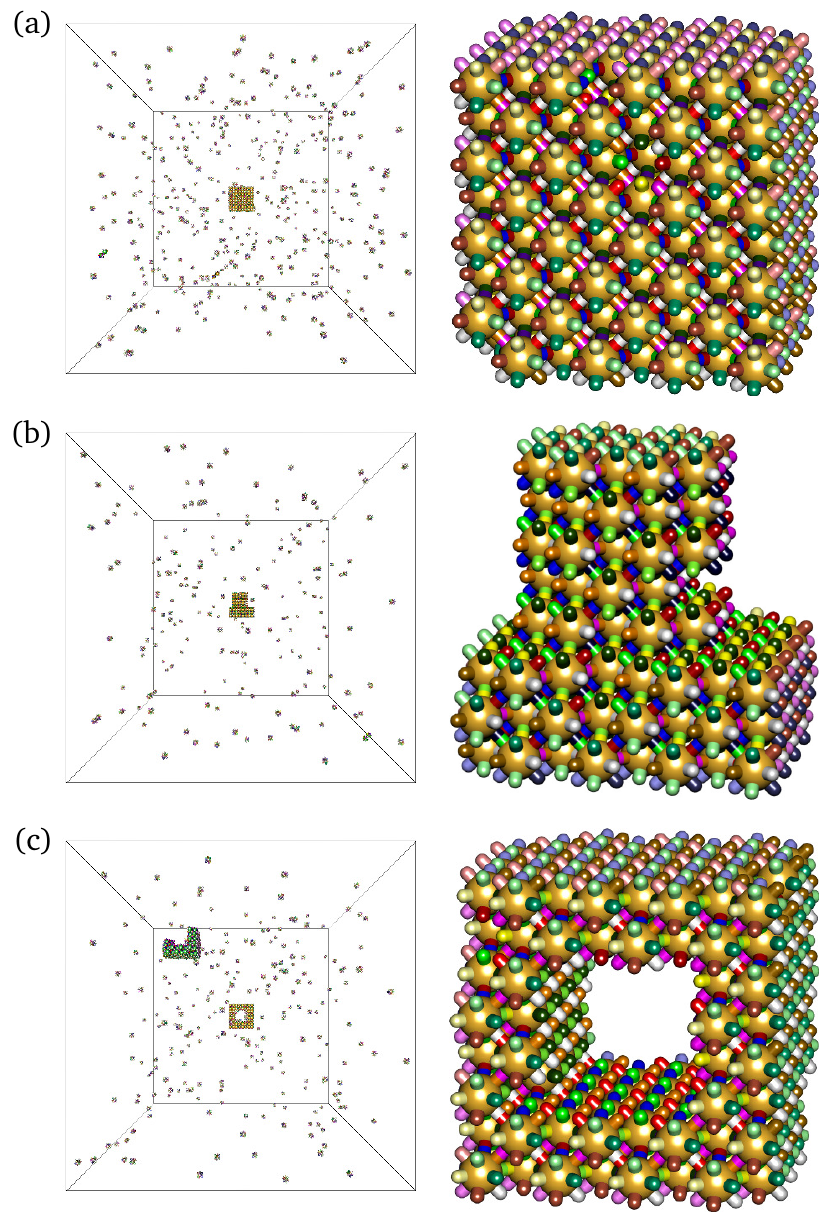}
\caption{Snapshots from grand-canonical ensemble simulations of 12-co-ordinate particles with a 68\,\% G-C content. $T=\SI{344}{\kelvin}$. The fugacity of all `desired' particle types is set to $z_\text{des}=2/(78a)^3$, where $78a$ is the length of the simulation box in lattice units. All simulation snapshots shown here were obtained from the same building blocks, but with setting the fugacity of particles not part of the `desired' structure to zero. In each case, the whole simulation box and a close-up of the largest cluster are shown. (a) Original cubic target structure. (b) `Top hat' structure. (c) Central cavity structure.}\label{fig-coordNum-grandCanonical-complex}
\end{figure}

One of the main advantages of the work on DNA bricks has been their modularity, in the sense that a large range of target structures have been assembled from essentially the same building blocks: the cubic target structures considered so far can be thought of as a `molecular canvas'.\cite{Ke2012, Wei2012} It is possible in experiment to construct more intricate structures simply by not including the undesired bricks from the assembly pot, although in practice, poly-T DNA strands were used at every non-bonded position to minimise undesired interactions. To verify that this modularity continues to be a feature of target structures with a higher co-ordination number, we have run grand canonical simulations with certain building blocks simply missing. This results in the self-assembly of more complex target structures, exactly as expected. We show two structures that have formed in such conditions in Fig.~\ref{fig-coordNum-grandCanonical-complex}: a `top hat' style structure and a cube with a cavity. The self-assembly of these target structures from a cubic canvas confirms that the modularity of the building blocks remains a feature in these high-co-ordination number structures.

Moreover, we have run simulations in which the target number density of the undesired building blocks is not set to zero, but rather to a finite but small number. In principle, one would expect that the undesired building blocks need not be completely absent from the reaction mixture, but must simply be vastly outnumbered by the correct building blocks. Our simulations suggest that this is indeed possible, but the fugacities (and hence the solution number densities) of the undesired particles must be set to very low values in order to form the target structure reproducibly. The precise value of the required fugacity depends on the environment of the undesired particles in the underlying canvas structure. For example, for structures with a co-ordination number of 12, if the target structure is a `top hat' (Fig.~\ref{fig-coordNum-grandCanonical-complex}(b)), most of the undesired particles are outlying particles with relatively few bonds connecting them to the remaining structure. It thus proves possible to form the desired target structure reliably when the undesired particle fugacities are set to approximately \SI{0.5}{\percent} of the desired particle fugacities of $z_\text{des}=2/(78a)^3$ (where $78a$ is the length of the simulation box in lattice units). Larger `undesired' fugacities result in considerable attachment of the undesired particles over time. However, if the target structure is the central cavity structure of Fig.~\ref{fig-coordNum-grandCanonical-complex}(c), most of the undesired particles are at the centre of the cubic canvas and any undesired bonding that does occur is rather stable; therefore an even lower concentration of undesired particles is required in order to be able to self-assemble the target structure robustly.\footnote{In addition, such a structure is considerably more difficult to nucleate than the full cube, since the nucleus that forms must be near the edges of the target structure and has, of necessity, fewer bonds and is thus less stable.}  Whilst in theory, designed structures can form in a modular way even when the solution concentration of undesired particle is non-zero, if the target structures are not passivated as they are in experiment (with a poly-T sequence assigned to outlying non-bonding portions of the single-stranded DNA), there is always the chance that at least some undesired particles will attach to the structure, either during growth or once the desired structure is already fully assembled. In this sense, the experimental strategy of passivating the outer surfaces appears to be very important and permits the desired structure to be assembled even in slightly unclean environments.

\section{Conclusions}

We have shown that, using a simplified computational model for addressable self-assembly, we are able to self-assemble structures with co-ordination numbers as high as 12. This was a somewhat unexpected result, because we had previously predicted that such structures will exhibit free-energy barriers to nucleation very different -- and less conducive to self-assembly -- from those previously determined for tetrahedrally co-ordinated structures. Our theoretical work suggested that the nucleation barriers would be less jagged in appearance and much more classical in shape. We predicted that this indicated that self-assembly would be considerably more challenging, because the supersaturation required for nucleation free-energy barriers to be surmountable would need to be greater: so great, we hypothesised, that competition from misassembled structures would dominate and it would be impossible for high co-ordination number structures to be assembled spontaneously in high yield.\cite{Jacobs2015b}

Indeed, the theoretical predictions we made about the free-energy barrier are borne out in simulations, but the hypothesis that such structures would be impossible to form is not. We have shown that the free-energy barriers do indeed become less jagged, the critical cluster size is considerably more temperature-dependent and it is more difficult to find mild conditions under which error-free self-assembly can occur. However, we have shown that despite this, it is still possible to find conditions under which the nucleation free-energy barrier is large enough that nucleation is rare, but sufficiently small that it can nonetheless sometimes occur, in conditions under which the stable structure lies along the pathway towards the formation of a fully assembled designed target structure. This is very good news, because it gives us some confidence that higher co-ordination number structures, which are expected to be of considerable interest in nanotechnology, may indeed be possible to assemble using only a simple protocol.

We have also shown that the design process is modular in much the same way as it is for DNA bricks and that the designed structures self-assemble reproducibly in computer simulations. However, it is necessary to qualify these successes of the simulation method. The computational model we have used to study these effects is very crude and neglects a number of aspects that are likely to be important in any experimental realisation. Notwithstanding the molecular-level mechanisms of DNA hybridisation that have been coarse-grained away, one particular limitation of the model we have used is that it is a lattice model, which over-constrains the geometry of the growing structures and favours their successful assembly. This geometric constraint may be a significant issue in experimental work, perhaps especially so if DNA multi-arm motifs rather than coated colloidal particles were used in the assembly process, as they are themselves not very stiff, and the resulting poor geometry of the growing cluster may significantly retard the growth process. Such additional geometric considerations may cause difficulties not only during the nucleation stage itself, where the additional loss of entropy of the monomers required to form a compact structure would likely increase the height of the free-energy barrier, but because of the time involved in the reorganisation of the monomer structure when bonding to the growing clusters, they may also reduce the ratio of the rate of cluster growth relative to cluster diffusion: this may make it more likely for different clusters in the system to meet and interact, frustrating their correct assembly. It would be useful in future work therefore to characterise more fully the effect of the cooling protocol on addressable self-assembly.

These considerations may mean that not all possible experimental approaches to many-component building blocks may result in successful self-assembly, and so experimental success is far from guaranteed. It is likely to be the case that an experimental realisation of such building blocks may involve a significant investment of time, effort and not least money. Nevertheless, since we have shown that high co-ordination number self-assembly is computationally feasible, this indicates that the underlying physics does not preclude such structures from self-assembling: we hope this will help to stimulate experimental efforts to achieve similar complexity.

\section*{Acknowledgements}
We thank William M.~Jacobs and David M.~Smith for helpful discussions. This work was supported by the Engineering and Physical Sciences Research Council [Programme Grant EP/I001352/1]. Research carried out in part at the Center for Functional Nanomaterials, Brookhaven National Laboratory, which is supported by the US Department of Energy, Office of Basic Energy Sciences, under Contract No.~DE-SC0012704. Supporting data are available at \href{https://www.repository.cam.ac.uk/handle/1810/250582}{https://www.repository.cam.ac.uk/handle/1810/250582}.

\providecommand*{\mcitethebibliography}{\thebibliography}
\csname @ifundefined\endcsname{endmcitethebibliography}
{\let\endmcitethebibliography\endthebibliography}{}

\end{document}